\documentclass[12pt]{iopart}
\usepackage{graphicx} % Include figure files
\usepackage{subfig}
\usepackage{bm}% bold math
\usepackage{color}

\begin{document}

\title{A local Fock-exchange potential in Kohn-Sham equations} 
\author{T.W. Hollins$^1$, S.J. Clark$^1$, K. Refson$^{2,3}$ and N.I. Gidopoulos$^1$\\
$^1$ Department of Physics, Durham University, South Road, Durham, DH1 3LE, United Kingdom.\\
$^2$Department of Physics, Royal Holloway, University of London, Egham TW20 0EX, United Kingdom.\\
$^3$ISIS Facility, Science and Technology Facilities Council, Rutherford Appleton Laboratory, Didcot OX11 0QX, United Kingdom.
}

\date{\today}

\begin{abstract}

We derive and employ a local potential to represent the Fock exchange operator in electronic single-particle equations.  
This local Fock-exchange (LFX) potential is very similar to the exact exchange (EXX) potential in density functional theory (DFT).
The practical software implementation of the two potentials (LFX and EXX) yields robust and accurate results for a variety of systems 
(semiconductors, transition metal oxides) where Hartree Fock and popular approximations of DFT typically fail. 
This includes examples traditionally considered qualitatively inaccessible to calculations that omit correlation.

\end{abstract}

%\pacs{31.15.E-, 31.10.+z, 31.15.xt, 71.15.-m}

\maketitle

\section{Introduction}

The exchange symmetry in quantum mechanics refers to the invariance of a quantum system when two identical particles 
exchange positions. 
For electronic systems in particular, this symmetry leads to Pauli's exclusion principle that makes necessary 
the use of antisymmetric, or fermionic, wave functions.

However, in the theory of electronic structure, the term ``exchange''
is often used in the narrower context of an effective description that
treats the interacting electrons as independent particles, assigning a
spin-orbital ($\phi_i$) to each one of them.  The $N$-electron system
is represented by a Slater determinant ($\Phi$) built on the set of
spin-orbitals $\{\phi_i\}$ and the electrons do not interact directly
with each other but each electron lies in the effective field of the
other $N-1$ electrons.  The two principal examples of such effective,
independent-particle descriptions are the Hartree-Fock (HF)
approximation \cite{szabo} and the Kohn-Sham (KS) scheme \cite{ks} in
density functional theory (DFT) \cite{hk}.
Based on the reference state $\Phi$ of the model, 
the definition of the exchange energy is given in the context of such an independent-particle theory,
%in terms of the spin-orbitals of $\Phi$,
\begin{equation} \label{one}
E_{\rm x} [ \Phi ] = - \frac{1}{2} \sum_\sigma \int\int \! d{\bf r} \, d{\bf r}' \,
\frac{ | \rho_\Phi^\sigma ({\bf r}, {\bf r}') |^2}{|{\bf r} - {\bf r}'| } .
\end{equation}
where $\rho^\sigma_{\Phi} ({\bf r}, {\bf r}')$ with $\sigma = \uparrow , \, \downarrow $
is the one-particle reduced density matrix of the reference Slater determinant $\Phi$.

Hence, the exchange energy is defined differently in HF theory, where $\Phi$ is the HF Slater determinant $\Phi_{\rm HF}$, from DFT where $\Phi$ is the 
 KS Slater determinant $\Phi_{\rm KS}$.
The two different definitions lead naturally to different representations of the exchange term (local in KS, nonlocal in HF) in the corresponding 
single-particle equations.

In HF the nonlocal Fock exchange term acts in a different way on the occupied and on the virtual orbitals and this asymmetry 
can lead to counterintuitive and unphysical behaviour of the HF orbitals and their energies.
For example, the virtual HF orbitals appear to be repelled by a charge of $N$ rather than $N-1$ 
electrons\cite{davidson_rmp}, a fact that has led to the interpretation of the virtual orbital energies as 
(negative) electron affinities, in an obvious extension of Koopmans' theorem \cite{koop}.
%This interpretation seems accidental, as the derivation of the HF equations does not justify interpreting the virtual HF orbitals 
%as orbitals of the $N+1$ electron system.
On the other hand, if we view that a virtual HF orbital represents an excitation of a ground state orbital, then the repulsion of the virtual orbital 
by a charge of $N$ rather than $N-1$ electrons becomes a qualitative error arising from the 
self-repulsion of the occupied orbital -- accommodating the electron before excitation -- with the virtual orbital 
hosting to the electron after excitation.
Ref.\cite{ghost} discusses a similar case of self-repulsion (``ghost-interaction'') in ensemble DFT for excited states.

The self-repulsion of the virtual orbitals and the asymmetry of the action of the Fock operator on occupied and virtual orbitals 
leads to too high single-particle excitation energies and to poor band-structures with too large band gaps. 
In metals, the density of states of the uniform electron gas is found to vanish at the Fermi energy\cite{ashcroft}, 
predicting wrongly the behaviour of an ideal metal to be almost insulating. (However, see Ref.~\cite{aan}.)
Another qualitative error of these spurious self-repulsions is that in unrestricted HF theory
there are no unfilled shells, i.e., the highest occupied and the lowest unoccupied orbitals 
are never degenerate, even in systems with odd number of electrons\cite{lieb}.
Finally, it seems counterintuitive that all the occupied HF orbitals turn out to have the same asymptotic decay\cite{handy}.

The asymmetry in the HF treatment of occupied and virtual orbitals is well known and there is significant
work to correct it, e.g. by making rotations in the Hilbert space of virtual orbitals\cite{davidson}. 
An elegant solution to this problem is to employ a common, local, multiplicative, single-particle exchange potential 
$\hat v_{\rm x} = v_{\rm x}({\bf r})$, 
that treats all orbitals, occupied and virtual, in a symmetric way.
Indeed, with the self-interaction-free ``exact exchange'' (EXX) potential in KS theory (see Ref.\cite{kummel_kronik} and references therein) 
%and the exchange optimised effective potential (xOEP)~\cite{nt0,sharp,talman,kummel_kronik}, 
single-particle properties, such as ionisation potentials\cite{kummel_kronik,gorling}, electron affinities\cite{kummel_kronik}, 
single-particle excitation energies and band structures\cite{kummel_kronik,g1,g2,us} are obtained accurately. 
We note that the EXX potential cannot be written explicitly in terms of the electron density, but it must be 
obtained indirectly from the density, by solving a Fredholm integral equation of the first kind, 
known as the equation for the optimized effective potential (OEP) method~\cite{sharp,talman,kummel_kronik}.  
In the terminology of KS theory, when correlation is omitted the EXX potential is also 
referred to as the exchange optimized effective potential (xOEP).

By comparison, in semi-local KS approximations (such as the local density approximation, LDA, 
and other related approximations) the cancellation of self-interactions is incomplete making the asymptotic decay of the
exchange potential too fast, thus leading to inferior single-particle properties. 
In fact, the enforcement of the correct asymptotic behaviour is sufficient to improve
considerably the accuracy of single-particle properties in these approximations~\cite{clda_nn}.

Given that the exchange energy and exchange potential are not measurable quantities, 
the question arises about their dependence on the reference state of the effective model. 
For example, %is this dependence random, or, 
are there physical limits where this dependence can be expected to be either strong or weak? 
This question will be investigated in the following, after obtaining the local 
exchange potential (``local Fock exchange'', LFX) corresponding 
optimally to HF's nonlocal Fock exchange term.

\section{The local Fock exchange potential}

The KS system is a virtual system of noninteracting electrons defined to have the same ground state 
charge density as the interacting system of interest.
The KS system is determined by solving single-particle (KS) equations featuring the KS potential, which 
forces the noninteracting electrons to have the required density. 

In DFT, the KS potential (functional derivative w.r.t. density of the noninteracting kinetic energy functional) 
is the sum of the electron-nuclear attraction, $v_{\rm en} ({\bf r})$, the Hartree potential 
$\int d{\bf r}' \rho({\bf r}') / | {\bf r} - {\bf r}' | $, 
and the exchange and correlation potential, $v_{\rm xc} ({\bf r})$; the latter is the functional derivative w.r.t. the density of
the exchange and correlation energy. 
In particular the xOEP is the functional derivative of the exact exchange energy, given by the 
Fock expression (\ref{one}) in terms of the KS orbitals.

Ref.~\cite{pra} gives an alternative (to DFT) and direct way to obtain ab initio approximations for the KS potential, 
by minimizing an appropriate energy difference ($T_\Psi [v]$, see the variational principle (4) in \cite{pra}).
We note that the formalism in \cite{pra} assumes some knowledge of the interacting ground state $\Psi$. 
Also, we point out that although the theoretical scheme developed in Ref. \cite{pra}, is based on wave function theory rather than DFT, 
it still results in the usual KS single-particle equations employing the (exact or approximate) KS potential.

In the following, to derive the local potential that best simulates the nonlocal Fock exchange term,
we follow Ref.~\cite{pra}, and use the HF ground state Slater determinant $\Phi_{\rm HF}$ in place of the 
interacting state $\Psi$. Then, we search among all effective Hamiltonians 
$H_v $ characterised by a local potential 
$v ({\bf r})$,
\begin{equation} \label{dyo}
H_v = \sum_{i=1}^N \left \{ - { \nabla_i ^2 \over 2} + v ({\bf r}_i) \right \}  , 
\end{equation}
for the one ($H_{v_{{\rm MP}0}}$) which adopts optimally %the HF ground state Slater determinant 
$\Phi_{\rm HF}$ as its own approximate ground state.
%Hamiltonian is $H_v = \sum_{i=1}^N [ - (1/2 ) \nabla_i ^2 + v ({\bf r}_i) ]$. 
%
The optimisation is achieved using the Rayleigh Ritz variational principle
and minimising, over all local potentials $v({\bf r})$, the non-negative energy difference,
\begin{equation} \label{vphf}
T_{\rm HF} [v] \ge 0 , \ {\rm with} \ 
T_{\rm HF} [v] \doteq \langle \Phi_{\rm HF} | H_v | \Phi_{\rm HF} \rangle - E_v .
\end{equation}
$E_v$ is the ground state energy of $H_v$. %the effective Hamiltonian. 
The functional derivative of $T_{\rm HF} [v]$ is equal to the difference of the HF density and the density of the effective system~\cite{pra},
\begin{equation} \label{fd}
\frac{\delta T_{\rm HF} [v] }{ \delta v ({\bf r})} = \rho_{\rm HF} ({\bf r}) - \rho_v ({\bf r}) .
\end{equation}
At the minimum, the two densities are equal and the optimal potential $v_{{\rm MP}0}$ has the same density as HF. %$\Phi_{\rm HF}$. 
The difference of $v_{{\rm MP}0}$ and the sum of the Hartree potential and the electron-nuclear attraction defines the 
local Fock-exchange (LFX) potential:
\begin{equation}
v_{\rm LFX} ({\bf r}) \doteq v_{{\rm MP}0} ({\bf r}) - v_{\rm en} ({\bf r}) - \int \! d{\bf r}' \, \frac{\rho_{\rm HF}({\bf r}') }{ | {\bf r} - {\bf r}' |} .
\end{equation}
Compared with the exchange optimised effective potential (xOEP) 
case, the functional derivative in (\ref{fd}) is easier to obtain, as it 
does not involve the calculation of the orbital shifts\cite{us,kummel_prl}.

The local potential with the HF density was studied in the past but also recently, in Ref.~\cite{rya,sta2}, 
where it was found that it provides an almost perfect approximation to xOEP, avoiding mathematical issues in 
the implementation of xOEP with finite basis sets.   

However, from our variational derivation we argue that LFX is actually more than a mere approximation to xOEP: 
both potentials are optimal in Rayleigh-Ritz energy minimizations where we search for 
the effective Hamiltonian $H_v$ that, either adopts the HF ground state $\Phi_{\rm HF}$ 
optimally as its own approximate ground state (LFX), or, whose ground state minimizes the HF
total energy (xOEP). In either case, if the restriction of a local potential were relaxed, the HF Hamiltonian 
would be the minimizing one. 
It follows that both the LFX potential and the xOEP simulate optimally 
the nonlocal Fock exchange term in the HF single-particle Hamiltonian.

In the following we demonstrate that the xOEP and the LFX potential share other characteristic properties associated 
in the literature with exchange. For example,
they both form zero-order terms in separate power series expansions of the 
KS potential, where the first order corrections vanish~\cite{nig}, as we show below.
%In DFT Perturbation Theory (PT) this property defines the exchange potential. 

At the variational derivation of $v_{{\rm MP}0}$ above, if in place of $\Phi_{\rm HF}$ we employed a finite-order expansion of
the interacting ground state using M{\o}ller-Plesset (MP) Perturbation Theory (PT) \cite{szabo}, %truncated at various finite orders 
and then at each order performed the same optimisation, 
we would generate a corresponding MP series expansion for the KS potential.
From Brillouin's theorem \cite{szabo}, singly excited Slater determinants 
do not couple directly with the HF ground state and so, 
the density of the MP ground state does not change from zero to first order.
Hence, in the MP expansion of the KS potential, the zero-order term is $v_{{\rm MP}0}$ and the first-order correction vanishes.
The same holds true for xOEP \cite{b1,gl2} in DFT PT \cite{gl}:
%According to G\"orling, Levy and Bartlett and coworkers, 
the sum of the xOEP, the Hartree potential and $v_{\rm en}$, 
form the local potential in the zero-order effective Hamiltonian $H_v$, where, 
if we switch on electronic repulsion %$V_{\rm ee} - \sum_i v({\bf r}_i)$, 
the electronic density will not change to first order.
This result was re-derived in Ref.~\cite{pra}, using a general version of the variational principle (\ref{vphf}).

Probably the most characteristic property of the exchange potential is that it is the functional derivative of the exchange energy 
(\ref{one}) and as a consequence xOEP satisfies the virial relation for exchange:
\begin{equation} \label{ve}
E_{\rm x} [\Phi_{\rm xOEP}]+ \int dr \, \rho({\bf r}) \, {\bf r} \! \cdot \! \nabla v_{\rm xOEP} ({\bf r}) = 0 .
\end{equation}
The LFX potential is not the functional derivative of the exchange energy, satisfying a different virial relation \cite{nig}:
\begin{equation}
 E_{\rm x} [\Phi_{\rm LFX}] + \int dr \, \rho({\bf r}) \, {\bf r} \cdot \nabla v_{\rm LFX} ({\bf r})+  E_{\rm c}^{\rm HF} + T_{\rm c}^{\rm HF} = 0 \, , 
\end{equation}
where, 
\begin{eqnarray}
T_{\rm c}^{\rm HF} &  \doteq &  \langle \Phi_{\rm HF} | T | \Phi_{\rm HF} \rangle - \langle \Phi_{\rm LFX} | T | \Phi_{\rm LFX} \rangle \\
E_{\rm c}^{\rm HF}  & \doteq  & \langle \Phi_{\rm HF} | T+ V_{\rm ee} | \Phi_{\rm HF} \rangle - \langle \Phi_{\rm LFX} | T+ V_{\rm ee} | \Phi_{\rm LFX} \rangle
\end{eqnarray}
$V_{\rm ee}, T$ are the interaction and kinetic energy operators.
%Sharing the same ground state density, $H_{v_{{\rm MP}0}}$ (at $\lambda=0$) 
%and the HF Hamiltonian (at $\lambda = 1$) can be connected by an adiabatic-connection path 
%\cite{langreth_perdew, gunnarsson_lundqvist,harris_jones}. The latter is represented by a 

Consider now the perturbation treatment of the $N$-electron HF Hamiltonian, 
with zero order $H_{v_{\rm MP0}}$ (Eq.~\ref{dyo}, with $v = v_{\rm MP0}$) and perturbation 
$ \hat K - \sum_i v_{\rm LFX} ({\bf r}_i) $, 
\begin{equation}
H_{\rm HF} = H_{v_{\rm MP0}} + \lambda \left( 
\hat K - \sum_i v_{\rm LFX} ({\bf r}_i) 
\right) ,
\end{equation}
where by $\hat K$ we denote the many-body, nonlocal Fock exchange term. 
The zero order Hamiltonian and the perturbation are composed of one-body operators (both $H_{\rm HF} $ and $H_{v_{\rm MP0}}$ 
have similar Slater determinant eigenstates) and hence the perturbation expansion in powers of $\lambda$ is expected to converge fast.

Since they share the same ground state density, $H_{v_{\rm MP0}}$ and $H_{\rm HF} $ can be connected with 
an adiabatic connection path along $\lambda$, for $0 \le \lambda \le 1$ \cite{langreth_perdew,gunnarsson_lundqvist,harris_jones}.
Then, following Levy and Perdew~\cite{lp} we obtain, $E_{\rm c}^{\rm HF} + T_{\rm c}^{\rm HF} = 0 $, to second order in $\lambda$,
%this rapidly convergent expansion, 
and consequently {\em the LFX potential satisfies the virial relation for exchange} (\ref{ve}) {\em to second order}  in this rapidly convergent expansion in $\lambda$.
This explains the nearly perfect exchange-virial results in Ref.~\cite{rya,sta2}.

%Finally, in the Sham Schl\"uter equation \cite{book_hardy}, when the one-particle Green's function 
%$G$ and self energy $\Sigma$ are given in the HF approximation,
%($G=G_{\rm HF}, \Sigma=\Sigma_{\rm HF}$), 
%the solution of the Sham Schl\"uter equation yields the potential $v_{{\rm MP}0}$ with the HF density.
%The simpler linearised form of the %Sham Schl\"uter 
%equation, 
%obtained by replacing $G_{\rm HF}$ with the $G_v$ (one-particle Green's function for effective local potential $v$), 
%reduces to the xOEP equation [cite Gross Dreizler]. 
%From this point of view xOEP is an approximation to the LFX potential.  
 
We conclude that the difference between the xOEP and the LFX potential is analogous to the 
difference in the definitions of the exact exchange energy in wave function theory and in DFT.
Although the two potentials are mathematically different, 
they describe the same physics and consequently we expect the results 
of calculations employing these two potentials to be similar, at least in systems where exchange dominates over correlation (weakly interacting).
%Our comparison of the two methods shall reveal that in systems characterised by stronger correlations, a 
%small disparity between the two exchange methods can appear, which is indicative of the strength of correlations.  

\section{Algorithm to construct the LFX potential}

We now discuss our practical implementation method to calculate the LFX potential. 
The variational principle, $T_{\rm HF}[v] \ge 0$, suggests the following steepest descent algorithm 
to determine the potential $v_{{\rm MP}0}$:
%\begin{itemize}
%\item

$\bullet$ Start with a trial potential $v$ that yields a ground state density $\rho_v$. 
%\item

$\bullet$  If $\rho_v \ne \rho_{\rm HF}$, correct the potential in the direction, 
\begin{equation} \label{ch}
 v ({\bf r})  \rightarrow v ({\bf r}) - 
 \epsilon %\int {dr \, \rho_{\rm HF} ({\bf r}') - \rho_{v} ({\bf r}') \over | {\bf r} - {\bf r}'| } 
\int  \! d{\bf r}' \, \frac{ \rho_{\rm HF} ({\bf r}') - \rho_{v} ({\bf r}') }{ | {\bf r} - {\bf r}' | }
\end{equation}
where $\epsilon  >0$ is a small, positive real number. 

%\item 
$\bullet$ Recalculate $ \rho_v $ for the corrected potential $ v(\mathbf{r}) $.

%\item 
$\bullet$ Iterate to convergence.

For $\rho_v \ne \rho_{\rm HF}$ and sufficiently small %to first order in 
$\epsilon$, the change of the potential (\ref{ch}) reduces the energy difference in (\ref{vphf}) %$\delta T_{\rm HF } = - 
by $\epsilon \, U_{\rho_{\rm HF}} [ v ]$, 
where $U_{\rho_{\rm HF}} [ v ]$ is the Coulomb energy of the density difference 
$\rho_{\rm HF} - \rho_v$:
%by $\epsilon \, Q_{\rho_{\rm HF}} [ v ]$, 
%where $Q_{\rho_{\rm HF}} [ v ]$ is the integrated-squared density difference: 
\begin{equation}
U_{\rho_{\rm HF}} [ v ] \doteq \int \! d {\bf r} d {\bf r}' \,
 \frac{ [\rho_{\rm HF} ({\bf r}) - \rho_{v} ({\bf r})  ] [ \rho_{\rm HF} ({\bf r}') - \rho_{v} ({\bf r}') ] }{ | {\bf r} - {\bf r}' | }  \ge 0
\end{equation}
%\begin{equation}
%Q_{\rho_{\rm HF}} [ v ] \doteq \int \!\! d {\bf r} \,
%[\rho_{\rm HF} ({\bf r}) - \rho_{v} ({\bf r})  ]^2  \ge 0 .
%\end{equation}
The algorithm stops only when the two densities become equal, 
within a small computational tolerance. 
In practice a numerical optimization method, such as conjugate gradient, is used to accelerate convergence.  

It can be shown that for small enough $\epsilon$, the change in (\ref{ch}) reduces also the Coulomb energy $U_{\rho_{\rm HF}} [ v ]$ in every iteration.
In fact, the algorithm is general and for any target density $\rho$, 
the minimisation using (\ref{ch}) of $U_{\rho} [v]$ (the Coulomb energy of the density difference $\rho - \rho_v $), 
can be employed to invert $\rho$ and obtain the minimising local potential with ground state density equal to $\rho$. %~\cite{fts}. 
For related algorithms, see Refs.~\cite{wuyang2003,peirs} and the discussion in \cite{baerends,sta3}.

The functional derivative, $\delta T_{\rm HF} [v] / \delta v ({\bf r})$, % = \rho_{\rm HF} ({\bf r}) - \rho_v ({\bf r}) $, 
represents a charge density (\ref{fd}) with zero net charge 
(as does the functional derivative, $\delta E [v] / \delta v ({\bf r})$, of any energy expression $E[v]$ that is a functional of a local potential $v({\bf r})$; 
the net charge, $\int d{\bf r} \, \delta E [v] / \delta v ({\bf r})$, vanishes because the potential is defined up to a constant). \\
The correction of the potential in (\ref{ch}) is in the opposite 
direction of the Coulomb potential of that charge density.
In Ref. \cite{us} we had employed a different algorithm, where the potential was corrected in the opposite direction of the functional derivative, 
$\delta E [v] / \delta v ({\bf r})$, rather than the Coulomb potential of the functional derivative (see Fig.~1 in Ref.~\cite{us}). 
Here, the analogous algorithm to Ref. \cite{us} would have been: $v ({\bf r} ) \rightarrow v ({\bf r}) - \epsilon [ \rho_{\rm HF} ({\bf r}) - \rho_v ({\bf r}) ]$.
We found that this algorithm too converges to the same LFX potential, 
but is less stable and suffers from slow convergence rate in regions of low density. 

\subsection{Implementation and Convergence}

The LFX potential has been implemented in the electronic-structure, plane-wave code 
CASTEP \cite{castep1,castep2}, using the algorithm described above. The procedure to calculate the 
LFX potential, requires first a HF calculation to determine the target density ($\rho_{\rm HF}$). 
The second ingredient is the initial trial potential $v$, for which we choose 
the LDA potential corresponding to the target density; 
we have tried other density-dependent potentials and they perform equally well. The rest of the LFX potential calculation proceeds iteratively: 
we solve the single-particle equations with the potential $v({\bf r})$ 
and obtain the occupied orbitals and their density $\rho_v$. 
The difference of the Coulomb potentials of the densities, $\rho_{\rm HF} - \rho_v$ (see
Eq. \ref{fd}), is used to correct the potential using (\ref{ch}) in a
line search, where in the latter case, $\epsilon$ is chosen to
minimise $U_{\rho_{\rm HF}}[v]$.  
%%%sjc
This procedure gives a down-hill direction allowing implementation of
a Fletcher-Reeved based conjugate gradients algorithm with a line
search based on a parabolic two-step fit.
%%%sjc
The iterative procedure is repeated
until both $U_{\rho_{\rm HF}}[v]$ and %$\Delta U_{\rho_{\rm HF}}[v]$ (
the change in $U_{\rho_{\rm HF}}[v]$ with each iteration become
comparable or smaller than the
%threshold values, $10^{-5}$eV and $10^{-7}$eV.  % respectively.
threshold values of $1 \mu$eV and $\sim 0.01 \mu$eV respectively.
%$10^{-5}$e$^2$ \AA$^{-3}$ and $10^{-7}$e$^2$\AA$^{-3}$. % respectively.

In plane-wave DFT implementations the orbitals, 
density, and potentials are represented on rectilinear grids. 
The Kohn-Sham orbitals are described 
within a sphere bounded by the cutoff wave vector, $G_{max}$, and the density and 
potentials are nonzero within a sphere of radius $2G_{max}$. 
The iterative optimisation scheme we developed 
is performed explicitly on these 
real space grids by direct variation, so that the effective basis used to represent 
$v(\mathbf{r})$ is the set of grid points 
$ \{ \mathbf{G} \} : | \mathbf{G} | \leq 2G_{max} $.

In the calculations that follow, the basis set size (plane-wave cutoff energy) and Brillouin-zone sampling were 
chosen so that total energy differences, evaluated in xOEP, were less than 2.5 meV/atom.

\subsection{Finite basis OEP errors and CEDA potential}

The implementation of the OEP method involves the expansion of the orbitals and the potential in finite basis sets. 
This procedure may introduce spurious oscillations in OEP, % (but not in LFX which is obtained differently),
caused by a discontinuity in the solution of the finite-basis OEP equations~\cite{pra_nn}.  
In practice, we find that with the large orbital basis sets we use, containing several thousands of plane-waves, this discontinuity is negligible.
(The discontinuity of finite-basis OEP \cite{pra_nn} is expected to diminish with increasing size of 
orbital basis~\cite{pra_nn,pra_comment,pra_reply}.)%,fts}. 
%
%The determination of the LFX potential does not hide a discontinuity and such errors are avoided. 
%See the supplementary material for more details.
%Details of this study will be published elsewhere.

%The implementation of the OEP method usually involves the expansion of the orbitals and the potential 
%in finite basis sets, a procedure that introduces errors~\cite{pra_nn}.
%
%
%We find that with the large plane-wave orbital basis sets we use, %which contain several thousands of plane-waves, 
%the discontinuity of the solution of finite-basis OEP equations, underlying such errors \cite{pra_nn}, is negligible~\cite{footnote}. 
%
%The determination of the LFX potential does not involve the density-density response function and such errors are avoided. 
%
%Details of this study will be published elsewhere.

Finally, we implemented the ``common energy denominator approximation'' (CEDA) to xOEP \cite{ceda}.
CEDA is equivalent \cite{melBulat} to the ``effective local potential'' \cite{elp} and to the 
``localised Hartree Fock'' potential \cite{lhf}.
These approximate xOEPs are related to the Krieger-Li-Iafrate (KLI) approximation \cite{kli}, 
since they are all based on the Uns\"old approximation \cite{ceda0}. 

\section{Results}

\begin{table}
\begin{indented}
\caption{Total energy differences $\Delta E$ (in eV) from the HF
  energy and Kohn Sham bandgaps (in eV) for xOEP, LFX and CEDA
  potentials. Last column gives experimental bandgap values; from
  Refs.~\cite{us,engel}. \label{etottab}}
\item[]
\begin{tabular} {  c | c c c | c c c | c }
\br
\multicolumn{1}{c}{ } & \multicolumn{3}{c}{$\Delta E$ } & \multicolumn{4}{c}{ Band Gap}\\ 
\multicolumn{1}{c}{ }  & xOEP & LFX & CEDA & xOEP & LFX & CEDA & Exp.\\ \mr
Ge & 0.432 & 0.441 & 0.724 & 0.91 & 0.91 & 0.39 & 0.79 \\
InN & 0.467 & 0.480 & 0.700 & 1.36 & 1.32 & 0.73 & 0.93 \\
Si & 0.213 & 0.213 & 0.299 & 1.18 & 1.18 & 0.71 & 1.16 \\
GaAs & 0.428 & 0.444 & 0.718 &1.89 & 1.85 & 1.00 & 1.52 \\
CdTe & 0.390 & 0.397 & 0.572 & 2.22 & 2.16 & 1.57 & 1.61 \\
ZnSe & 0.472 & 0.485 & 0.702 & 2.89 & 2.85 & 2.14 & 2.80 \\
GaN & 0.401 & 0.416 & 0.652 & 3.29 & 3.27 & 2.65 & 3.39 \\
ZnO & 0.381 & 0.391 & 0.539 & 3.49 & 3.41 & 2.88 & 3.43 \\
C & 0.159 & 0.160 & 0.224 & 4.77 & 4.78 & 4.25 & 5.47 \\
CaO & 0.258 & 0.264 & 0.546 & 6.08 & 5.93 & 4.73 & 8.97 \\
NaCl & 0.050 & 0.050 & 0.112 & 6.28 & 6.23 & 5.47 & 7.09 \\
 & & & & & & & \\
FeO & 1.438 & 1.595 & 2.994 & 1.21 & 0.72 & 0.36 & 2.4\\
CoO & 1.595 & 1.698 & 3.193 & 2.26 & 1.95 & 1.11 & 2.5\\
MnO & 0.823 & 0.951 & 1.833 & 3.85 & 3.30 & 3.36 & 3.9\\
NiO & 1.647 & 1.717 & 3.403 & 3.93 & 3.74 & 2.72 & 4.0\\
\br
\end{tabular}
\end{indented}
\end{table}

The differences in total energies, from the HF total energy, for the
xOEP, LFX and CEDA potentials are shown in Table \ref{etottab}, for a
selection of semiconductors, insulators and anti-ferromagnetic
transition metal monoxides (TMOs). In every case we used the
HF total energy expression in terms of the occupied orbitals of the
corresponding potential.  Also included in Table \ref{etottab} are the
Kohn-Sham band-gaps for each material, calculated as the difference
between the conduction band minimum and valence band maximum.
The values for the experimental bandgaps for the
semiconductors and insulators are from Ref. \cite{us} and references
therein. The experimental bandgaps and magnetic moments
(Tables~\ref{etottab}, \ref{magmotab}) for the TMOs are from Ref. \cite{engel} and references therein.

The calculations on the semiconductors and insulators used
experimental lattice constants in the zincblende structure for Ge,
Si, GaAs, CdTe, ZnSe, C, the wurtzite structure for InN, GaN and ZnO
and the rocksalt structure for CaO and NaCl.

\begin{figure}
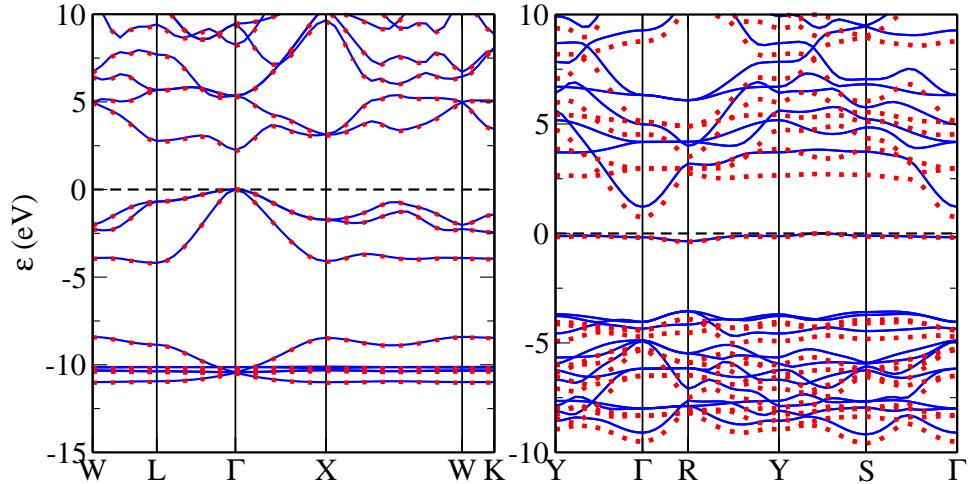

\begin{indented}
\caption{Band structures of CdTe (left) and FeO (right), blue solid
  lines are for xOEP and red dotted lines are for LFX. The blue and
  red lines are indistinguishable in CdTe but differ in FeO.  The
  discrepancy indicates stronger correlations.\label{BS-Plots}}
\item[]\subfloat{\includegraphics[scale=0.450, trim = 0mm 0mm 0mm 15mm,
    clip]{CdTe.eps}} \subfloat{\includegraphics[scale=0.450, trim =
    10mm 0mm 0mm 15mm, clip]{FeO.eps}}
\end{indented}
\end{figure}

The HF energy is the lowest among all the methods due to the greater
variational freedom of the orbitals compared to the local potential.
The total energies corresponding to the xOEP and LFX potentials are
higher by 0.5 eV and within 15 meV of each other, with the largest
difference observed in GaN. Of these the LFX energy is slightly
higher, in accord with the definition of xOEP as the minimum among
effective schemes with a local potential.  As expected, the CEDA total
energies are significantly higher.  Band structures generated using the
xOEP and LFX potentials are also very similar. A typical example,
CdTe, is plotted in Figure \ref{BS-Plots}, where the difference is
undetectable.  For materials not considered ``highly correlated'' the
largest difference of 0.15 eV for CaO would be barely visible on the
same scale. CEDA bandgaps are systematically smaller by between a few
tenths of eV and 1 eV.  Such close similarities support our contention that
the xOEP and LFX potentials capture equally well the physics of the
exchange term in the single-particle equations.

Our results agree well with EXX/xOEP results in the literature, showing that KS bandgaps 
(without a discontinuity correction) predict the fundamental bandgaps of semiconductors 
and insulators rather accurately, but with the quality of the agreement generally deteriorating with 
increasing width of the insulator bandgap~\cite{g1,g2,exxgross,kummel_kronik}.  

\subsection{Transition metal oxides}

Calculations for the TMOs were performed in the
experimental rocksalt structure without any rhombohedral
distortion. The simulation cell was the primitive rhombohedral cell of
the AFM II magnetic structure, which is consistent with
antiferromagnetic order between alternating cubic (111) planes.  All
four methods (HF, xOEP, LFX, CEDA) predict the four materials to be
antiferromagentic insulators.  The xOEP total energy is slightly below
the LFX total energy, differing by at most 0.16eV for FeO (whose xOEP
and LFX band structure is in Figure \ref{BS-Plots}); both of these
methods give energies very close to the HF minimum.  As shown in
Tables \ref{etottab}, \ref{magmotab}, the exchange-only Kohn-Sham
bandgaps and the calculated magnetic moments are close to the
experimental values.

%These oxides are classified as strongly-correlated materials, and in
%many studies their insulating nature has been characterised as Mott,
%or charge transfer, where electronic correlations play the dominant
%role.  
%Such type of strongly correlated insulating state is supposedly
%beyond the independent particle picture, which is expected to always
%give a metallic state.

For the TMOs studied here, antiferromagnetism allows the opening of a gap in the single particle
spectrum, although semi-local approximations (LDA/GGA) predict a
gap that is too small or zero. The better treatment of
self-interaction in hybrid functionals\cite{Hybrid,Clark}, the
Perdew-Zunger SIC\cite{PZ-SIC}, LDA$+$U~\cite{LDA+U}, 
or previous EXX calculations\cite{engel} all yield improved agreement with
experiment. 
%Notably, all of these examples included some form of
%correlation. We are unaware of a previous calculation that
%omitted correlation and yet gave reasonable results for the
%band gaps and magnetic moments.

Our exchange-only results in TMOs are in good agreement with EXX results 
by Engel\cite{engel} which included LDA correlation. 
%Engel also found that in the KLI approximation
%FeO is metallic. By contrast our CEDA calculations, which used a
%related but not identical approximation, predict this state to be
%insulating but with a small band-gap.
%
The close agreement between our exchange-only results and
experiment suggests that the well-known failure of the semi-local
LDA and GGA approximations is mainly due to the incomplete description of
exchange rather than the strength of correlation \cite{blugel}. 
%
%Our work explains the apparent paradox that the description of the 
%strongly correlated TMOs is so
%improved by the more accurate  formulations of exchange over LDA. %compared with LDA.  
%

However, it is well-known that the fundamental band-gap is not equal to the KS
band-gap, and that the exchange and correlation discontinuity
$\Delta_{xc}$ must be added in order to obtain the correct value. 
The magnitude of $\Delta_{xc}$ will decide if the accuracy of our
exchange only results for the materials presented here (especially the TMOs) is a coincidence:
if it is small, the description of these systems will be physically sound. 
In fact, if the intuition for the smallness of $\Delta_{xc}$ at xOEP or LFX is verified, 
then, correlation in these results is not completely ignored, since it is included in the value of $\Delta_{xc}$.

\begin{table}
\begin{indented}
\caption{Magnetic moments (in $\mu_B$) for transition metal monoxides
  for xOEP, LFX and CEDA. Experimental values from
  Ref. \cite{engel}.\label{magmotab}}
\item[]\begin{tabular} {  c | c c c | c }
\br
\multicolumn{1}{c}{ } & xOEP & LFX & CEDA & Exp.\\
\mr
FeO & 4.00 & 3.98 & 3.98 & 3.32, 4.2\color{white}{0} \\
CoO & 2.98 & 2.94 & 2.90 & 3.35, 3.98\\
MnO & 5.06 & 5.10 & 4.98 & 4.58, 4.79\\
NiO & 1.92 & 1.90 & 1.80 & 1.64, 1.90\\
\br
\end{tabular}
\end{indented}
\end{table}

\section{Discussion}

To conclude, we have presented a thorough study and comparison of two very similar but mathematically distinct local, single-particle, exchange potentials.
%
%We emphasize that the definition of the exchange energy depends on the
%reference state of the independent particle scheme.  
%The same holds
%true for the definition of the correlation energy, and any difference
%in the exchange energies will correspond to the opposite difference in
%the correlation energies.  
%Hence, from the point of view of formal
%DFT the LFX potential, defined differently from xOEP, contains a
%component of the correlation potential \cite{rya}.  
%
%
%Another corollary is that in
%In the absence of correlation, the two definitions
%for the exchange energy are expected to converge to the same result.
%
%Whether the two equivalent potentials (LFX and xOEP) should be expected to give the same or very similar results depends 
%on the particular system under study and rests on the importance of correlation for that system.
%%
%The two methods can be expected to give very similar results, at least in systems where the exchange energy dominates 
%over correlation, since in these systems correlation can be approximately ignored.

%\subsection{Conclusions}
We applied our methods to a variety of systems, ranging from semiconductors, to wide-band insulators and to TMOs.
For most systems, the results from the two calculations were very
similar and almost indistinguishable. The main message of our paper is the robustness of our results,
which bolsters confidence on the computational/numerical aspect of the
results, since two entirely different calculations of the same underlying physical 
quantity turn out to agree in the end~\cite{science}.

The results of the common energy denominator approximation (CEDA) demonstrate the inferiority of the approximate treatment of exchange, 
compared with the more accurate treatment afforded by xOEP and LFX.

%This
%becomes even more pertinent with concerns over the last decade that
%the finite basis implementation of the OEP method is pathological.
%With the plane-wave basis used in our code, we did not
%encounter any issues with finite basis errors.

The larger differences between LFX and xOEP in some systems,
especially the TMOs, suggests that correlation
plays a more significant role in those systems.  We argue that the
disparity between LFX and xOEP can be taken as a measure of the
strength of correlation, as in principle, there is no {\em a priori} guarantee
that xOEP and LFX should give the same result.
Only in materials where the
XC energy is dominated by the exchange term would we
expect the two methods to agree.

%\newpage

\ack
T.W.H. acknowledges the Engineering and Physical Sciences Research Council (EPSRC) for financial support, the UK national supercomputing 
facility (Archer), Durham HPC (Hamilton),
the facilities of N8 HPC, provided and funded by the N8 consortium and EPSRC (Grant No.EP/K000225/1) and finally the UK Car-Parrinello 
Consortium for support under Grant No. EP/F037481/1.

\end{document}